\documentclass[aps, prl,10pt,notitlepage,nofootinbib,superscriptaddress,showkeys,twocolumn]{revtex4-1}
\usepackage{amstext,amsmath,amssymb,amsfonts,bbm, amsthm}
\usepackage[latin1]{inputenc}
\usepackage{fancyhdr}
\usepackage{graphicx}
\usepackage{hyperref}
\usepackage{tocvsec2}
\usepackage{subfigure}

\usepackage{color}

\def\beq{\begin{equation}}
\def\be{\begin{equation}}
\def\ee{\end{equation}}
\def\bes{\begin{eqnarray}}
\def\ees{\end{eqnarray}}

\DeclareMathOperator{\tr}{tr}
\DeclareMathOperator{\im}{Im}

\DeclareMathOperator{\Tor}{Tor}
\DeclareMathOperator{\Vol}{Vol}

\renewcommand{\ker}{\operatorname{Ker}}

\newcommand{\alg}{\mathfrak{g}}

\def\f{\frac}

\def\pp{\partial}

\def\calA{{\mathcal A}}

\def\calF{{\mathcal F}}

\def\calM{{\mathcal M}}

\begin{document}
%%%%%%%%%%%%%%%%%%%%%%%%%%%%%%%%%%%%%%%%%%%%%%%%%%%

\title{\large \bf Gauge symmetries in spinfoam gravity: the case for ``cellular quantization''}
%fixing in discrete BF theory: the case for ``cellular quantization''}

\author{{Valentin Bonzom}}\email{vbonzom@perimeterinstitute.ca}
\affiliation{Perimeter Institute for Theoretical Physics, 31 Caroline St. N, ON N2L 2Y5, Waterloo, Canada}

\author{{Matteo Smerlak}}\email{smerlak@aei.mpg.de}
\affiliation{Max-Planck-Institut f\"ur Gravitationsphysik, Am M\"uhlenberg 1, D-14476 Golm, Germany}

\date{\small\today}

%%%%%%%%%%%%%%%%%%%%%%%%%%%%%%%%%%%%%
\begin{abstract}\noindent
The spinfoam approach to quantum gravity rests on a ``quantization'' of BF theory using $2$-complexes and group representations. We explain why, in dimension three and higher, this ``spinfoam quantization'' must be amended to be made consistent with the gauge symmetries of discrete BF theory. We discuss a suitable generalization, called ``cellular quantization'', which $(1)$ is finite, $(2)$ produces a topological invariant, $(3)$ matches with the properties of the continuum BF theory, $(4)$ corresponds to its loop quantization. These results significantly clarify the foundations -- and limitations -- of the spinfoam formalism, and open the path to understanding, in a discrete setting, the symmetry-breaking which reduces BF theory to gravity.

%The topological BF theory is defined in a discrete setting in any dimension. Agreement with the functional integral of the continuum is found upon identifying the combinatorial (Reidemeister) and analytic (Ray-Singer) torsions. The main steps are: the formulation of the on-shell reducible gauge symmetries, and the use of Schwarz's resolvent. This extends previous works in two directions: (i) a full gauge-fixing of the symmetries, (ii) the loop quantization in arbitrary dimension. We compare it with the spin foam proposal which is based on 2-complexes and argue that it should be amended to account for the degrees of freedom of quantum BF theory.

\end{abstract}
%%%%%%%%%%%%%%%%%%%%%%%%%%%%%%%%%%%%%%
\maketitle

\paragraph{Introduction.}

Since it was first advocated by Baez \cite{Baez1994a}, Reisenberger \cite{Reisenberger1994} and Rovelli \cite{Reisenberger1996}, the \emph{spinfoam approach} to quantum gravity has attracted considerable interest, resulting in over a hundred papers published on the topic every year. In a nutshell, the idea underlying this activity is that a ``spacetime-covariant, Feynman-style'' sum-over-histories formulation of background-independent field theories exists in the form of a \emph{weighted sum over $2$-dimensional cell complexes} \cite{Reisenberger1996}. This approach is believed to provide a successful quantization of the topological BF theory \cite{Freidel1998a,Baez:1999sr}, in the form of the Ponzano-Regge \cite{PR} and Ooguri \cite{Ooguri:1992eb} models (in three and four dimensions respectively), and work is underway to adapt it to general relativity, understood as ``BF theory with extra constraints'' \cite{Engle:2007uq}. Standard reviews of the spinfoam formalism are \cite{Baez:1999sr,Perez:2003vx}; the state of the art is presented in \cite{Rovelli2011c}.

In spite of strong efforts and promising results \cite{Freidel2002,Freidel2004, noui-perez, Barrett:2009ys, Baratin:2010wi, Baratin:2011tg,Dittrich:2012qb,perezrovelli} (and more references in \cite{Dittrich:2012qb}), %\footnote{Notably by Freidel, Barrett, Dittrich and their collaborators.} 
several outstanding problems with the Ponzano-Regge and Ooguri (PRO) models have remained open so far. We may list them as follows.
\begin{enumerate}
\item
\emph{Bubble divergences}. The original PRO partition functions are in general divergent. How should one regularize them?
\item
\emph{Topological invariance}. The PRO partition functions are formally invariant under changes of triangulations, up to divergent factors. How can one turn them into finite topological invariants?
\item
\emph{Relationship to the canonical theory}. The connection between the Ponzano-Regge model and loop quantum gravity in $3$ dimensions was established in \cite{noui-perez}. Can this connection be extended to $4$ dimensions and higher?
\item
\emph{Relationship to the continuum theory}. BF theory was quantized in the continuum in \cite{Blau1991, blau-thompson-torsion}, and was showed to be related to the Ray-Singer torsion. Are the PRO models similarly related to torsion? (See \cite{Barrett:2009ys} for a positive answer in certain three-dimensional cases.)
\item
\emph{Diffeomorphism symmetry}. Both the continuum BF action and the Einstein-Hilbert action are diffeomorphism-invariant. What is the fate of this symmetry in the PRO models?
\end{enumerate}

Mostly thanks to the work of Freidel \emph{et al.} \cite{Freidel2002,Freidel2004}, it has become clear that all five problems are related to the issue of \emph{identifying the BF shift symmetry in a discrete setting} and gauge-fixing it. No complete solution to this issue, however, has been proposed in the literature. The purpose of this letter is to argue that there is a good reason for this: when dealing with 2-complexes only, as in the spinfoam formalism, there is no shift symmetry. To identify this symmetry, one must instead resort to an extension of the spinfoam formalism including higher-dimensional cells. This realization paves the way to what we call \emph{cellular quantization}. This cellular quantization solves problems 1 to 4, and sheds interesting new light on problem 5.

%no satisfactory gauge-fixing of the BF shift symmetry can be found within the spinfoam formalism itself, i.e. dealing with $2$-complexes only.

%They do not implement a proper gauge-fixing of the BF shift symmetry, and for this reason they fail to reproduce the features of quantum BF theory established by Horowitz \cite{Horowitz1989a}, Blau and Thomson \cite{Blau1991}: topological invariance of the partition function and relationship to Ray-Singer torsion. This issue manifests itself in the form of ``bubble divergences'' in the Ponzano-Regge and Ooguri partition functions, and spoils the alledged relationship between spinfoam sums and topological quantum field theory.

The letter is organized as follows. We start by reviewing the basic properties of the continuum BF theory, emphasizing its gauge symmetries and relationship to analytic torsion. We then describe the ``spinfoam quantization'' of BF theory, as described e.g. in Baez's reference paper \cite{Baez:1999sr}. We show how to identify the gauge symmetries in a discrete setting and perform a quantization which does preserve the topological features of the continuum theory. Finally we establish that this cellular quantization corresponds to the loop canonical quantization. %An appendix contains an explicit example showing that a ``spinfoam-like'' regularization of the Ponzano-Regge-Ooguri partition functions, relying solely on $2$-dimensional structures, fails to produce a topological invariant.

Let us also mention that more details on our results will be given in a separate paper \cite{discreteBF}; in particular, an explicit proof of the breakdown of topological invariance of the PRO models regularized on $2$-complexes will appear there.

\paragraph{Continuum BF theory.}

BF theory was introduced by Horowitz \cite{Horowitz1989a} and Blau and Thompson \cite{Blau1991} as an exactly soluble diffeomorphism invariant theory, illustrating the connection between quantum gauge systems and manifold topology previously discovered by Schwarz \cite{Schwarz1978}. Defined in terms of a gauge field (or gauge connection) $A$ and a $\alg$-valued $(d-2)$-form $B$ on spacetime $M$, where $d=\dim M$ and $\alg$ is the Lie algebra of the gauge group $G$, its classical action reads
\be\label{bfaction}
S_{\textrm{BF}}(B,A)=\int_{M}\langle B\wedge F(A)\rangle.
\ee
Here $F(A)$ is the field strength of $A$ and the bracket denotes a non-degenerate symmetric bilinear form in $\alg$, typically the Killing form when $\alg$ is semisimple. The corresponding field equations are $F(A)=0$ and $D_{A}^{k-2}B=0$, with $D^{k}_{A}$ the covariant exterior derivative associated to $A$ acting on $\alg$-valued $k$-forms.\footnote{Note that $D_{A}^{k}$ is not the $k$-fold of composition of the covariant exterior derivative with itself.}

In addition to the usual gauge symmetry of a gauge field, the action \eqref{bfaction} is invariant under the \emph{shift symmetry}
\be\label{shiftsymmetry}
B\mapsto B+D_{A}^{d-3}\lambda_{d-3},
\ee
where $\lambda_{d-3}\in\Omega^{d-3}(M,\alg)$ is any $\alg$-valued $(d-3)$-form. When $d\geq 4$, this gauge symmetry is \emph{on-shell reducible}: given a \emph{flat} connection $\phi$, i.e. one such that $F(\phi)=0$, the map $\lambda_{d-3}\mapsto D^{d-3}_{\phi}\lambda_{d-3}$ is many-to-one. This is to say that the gauge modes $\lambda_{d-3}$ (the ``ghosts'') themselves possess a gauge symmetry, namely
\be
\lambda_{d-3}\mapsto\lambda_{d-3}+D^{d-4}_{\phi}\lambda_{d-4},
\ee
with $\lambda_{d-4}\in\Omega^{d-4}(M,\alg)$ representing ``ghosts for ghosts''. In turn, these new variables themselves may have a shift symmetry, and so on.

This structure naturally fits in the so-called \emph{twisted de Rham complex}
\be\label{twistedcomplex}
0\rightarrow \Omega^{0}(M,\alg)\xrightarrow{D^{0}_{\phi}}\dotsm\xrightarrow{D^{d-1}_{\phi}}\Omega^{d}(M,\alg)\rightarrow0.
\ee
In this cochain complex, the coboundary maps are the covariant exterior derivative $D^{k}_{\phi}$ and the $k$-cochains are elements of $\Omega^{k}(M,\alg)$, viz. $(d-2-k)$-stage ghosts. Also note that, for a given flat connection $\phi$, the space of solutions of the field equation for $B$ is the cohomology space $H_{\phi}^{d-2}$ derived from \eqref{twistedcomplex}.

The path-integral quantization of BF theory requires the gauge-fixing of this shift symmetry. This can be achieved by means of the \emph{resolvent method}, a generalization of the Faddeev-Popov trick to reducible gauge symmetries devised by Schwarz \cite{Schwarz1978}. Starting from the formal, pre-gauge-fixing expression
\be\label{unfixedBF}
Z_{\textrm{BF}}=\int DA\int DB\,\, e^{iS_{\textrm{BF}}(B,A)},
\ee
the resolvent method consists in extracting the ``volume'' of the space $\im D_{A}^{d-3}$ arising in \eqref{shiftsymmetry} by means of the complex \eqref{twistedcomplex}. This method will be outlined below, when we apply it to gauge-fix the discrete counterpart of BF theory. For now, let us simply state the result of this procedure in the continuum: the gauge-fixed partition function $Z'_{\textrm{BF}}$ can be written as a sum over the \emph{moduli space $\calM$ of flat connections} on $P$, with a summand given by the analytic torsion of the complex \eqref{twistedcomplex}, viz.
\be
\Tor_{[\phi]}=\prod_{j=0}^{d-1}\det\Big((D^{j}_{\phi})^{\dagger}D^{j}_{\phi}\Big)^{(-1)^{j}/2}.
\ee
Here $[\phi]$ denotes the gauge equivalence class of a flat connection $\phi $, and the dagger denotes the adjoint with respect to arbitrary inner products in the spaces $\Omega^{k}(M,\alg)$; the Ray-Singer torsion is independent on these inner products. In particular, $Z'_{\textrm{BF}}$ is a topological invariant of $M$ \cite{Blau1991}.\footnote{Strictly speaking, these results hold when the twisted de Rham complex is acyclic (i.e. has vanishing cohomology) for all flat connections $\phi$, implying that the moduli space of flat connections is discrete. An extension to the general case is discussed in \cite{Blau1991}; see also \cite{Adams}.} 

The torsion also provides the measure for transition amplitudes and for the inner product between boundary wave-functions. Assume that $M$ has two disconnected boundaries $N_1, N_2$. Wave-functions are square-integrable functions over the moduli space of flat gauge fields on $N_{1,2}$. The transition from an initial state $\Psi_1$ to a final state $\Psi_2$ through $M$ reads
\be \label{transition}
\langle \Psi_2\vert Z'_{\rm BF}\vert \Psi_1\rangle = \sum_{[\phi]\in\calM} \Psi_2^*([\phi])\,\Tor_{[\phi]}\,\Psi_1([\phi]).
\ee
%the resolvent method consists in extracting the ``volume'' of the gauge group $\ker D_{A}^{d-3}$ using the cochain complex \eqref{twistedcomplex}.\footnote{More precisely, the resolvent method works with any such cochain complex; the twised de Rham complex is the canonical one in this setting.} To make sense of this notion of ``volume'', pick an arbitrary metric on $M$, and lift it to an inner product in the cochain spaces $\Omega^{k}$. Assume furthermore that all the cohomology spaces of \eqref{twistedcomplex} vanish. Then we have in particular $\ker D_{A}^{d-4}=\im D_{A}^{d-4}$, hence (dropping the reference to $M$ and $P$)
%\be
%\Vol(\ker D_{A}^{d-3})=\det \Big((D_{A}^{d-4})^{\dagger}D_{A}^{d-4}\Big)^{1/2}\f{\Vol(\Omega^{d-4})}{\Vol(\ker D_{A}^{d-4})}.
%\ee
%Iterating this relation, we obtain
%\be
%\Vol(\ker D_{A}^{d-3})=\prod_{j=4}^{d-4}\det \Big((D_{A}^{d-j})^{\dagger}D_{A}^{d-j}\Big)^{(-1)^{j}/2}\Vol(\Omega^{d-j})^{(-1)^{j}}
%\ee
%Now, let us \emph{pretend} that the cochain spaces $\Omega^{k}$ have unit volume: this is the meaning of the expression ``dividing by an infinite volume'' underlying the gauge-fixing procedure. Then we find that \eqref{unfixedBF} should be replaced by
%\be
%g
%\ee

From our perspective, the moral of this review is that, if classical BF theory can be thought of as a theory of connections and $(d-2)$-forms, \emph{quantum BF theory} on the other hand involves the entire twisted de Rham complex \eqref{twistedcomplex}, with forms of all degrees.

\paragraph{``Spinfoam quantization''.}

Let us now describe the ``spinfoam quantization'' of BF theory, as presented e.g. in \cite{Baez:1999sr}. Assume that $G$ is compact, and that $M$ is equipped with a triangulation $\Delta$ and its dual cell complex $K$. Define a \emph{discrete connection} on $\Delta$ as an assignment of an element $g_{e}$ of the gauge group $G$ to each edge ($1$-cell) $e$ of $K$. Then for each face ($2$-cell) $f$ of $K$, define the holonomy $H_{f}$ along $f$ as the ordered product of $g_{e}$ attached to the edges on the boundary of $f$. The set of group elements $H=(H_{f})_{f}$ is the discrete analogue of the field strength $F(A)$ in the continuum.

Now, consider again the formal expression \eqref{unfixedBF}, and ``integrate over the $B$-field''. This gives
\be
Z_{\textrm{BF}}=\int DA\,\,\delta\big(F(A)\big)
\ee
with a functional delta function implementing the flatness of the connection. Thanks to the discretization on $K$, the formal measure $DA$ can be defined by means of the Haar measure $dg$ on $G$, and we can set
\be\label{discretepartition}
Z_{\textrm{BF}}=\int_{G^{E}}\prod_{e}dg_{e}\ \prod_{f}\delta(H_{f}).
\ee
Now, let us expand the Dirac delta on $G$ in characters,
\be\label{delta}
\delta(g)=\sum_{j}\dim(j)\tr D^{j}(g),
\ee
where $j$ ranges over the equivalence classes of unitary irreps $D^{j}(g)$ of $G$, and recall the identity
\be\label{proj}
\int_{G} dg\, \bigotimes_{l=1}^{d} D^{j_{l}}(g)=\sum_{\iota}\vert\iota\rangle\langle\iota\vert
\ee
for the projector on the $G$-invariant subspace of the tensor representation $\bigotimes j_{l}$, of which the ``intertwiners'' $\iota$ spans an orthonormal basis. Plugging \eqref{delta} and \eqref{proj} into \eqref{discretepartition} then gives, after some easy algebra,
\be\label{PRO}
Z_{\textrm{BF}}=\sum_{(j_{f})}\prod_{f}\dim(j_{f})\prod_{v}\{N_{d}j\}.
\ee
Here $N_{d}=3(d+1)(d-2)/2$, and $\{N_{d}j\}$ is the Wigner $N_{d}j$ symbol. This expression defines the Ponzano-Regge ($d=3$) and Ooguri ($d=4$) models. 

Unfortunately, \eqref{PRO} is known to be ill-defined in general; when the sum over representations in \eqref{PRO} is truncated to a finite value $\Lambda$, the sum diverges as $\Lambda\rightarrow\infty$. This phenomenon has been coined ``bubble divergences'' \cite{Perez:2000fs}, and was interpreted as an ``infrared effect'' \cite{Baez:1999sr,Perez:2000fs}. The connection between these divergences and the BF shift symmetry was understood in $3$ dimensions by Freidel and Louapre \cite{Freidel2002}, and a ``gauge-fixing'' scheme consisting in removing certain faces of $K$ was proposed \cite{Freidel2002,Freidel2004}. For non-trivial topologies, however, this scheme turned out to fail turning \eqref{PRO} into a finite number \cite{springerlink:10.1007/s00023-011-0127-y}.

It should be clear from the above discussion that the spinfoam scheme, which only relies on the $2$-skeleton of $K$, does not implement any gauge-fixing of the discrete shift symmetry; it simply amounts to a rewriting of the \emph{unfixed} partition fonction \eqref{unfixedBF}. This is consistent when $d=2$, in which case BF theory is nothing but the zero-coupling limit of Yang-Mills theory; but it is \emph{inconsistent} when $d\geq3$, as the gauge redundancy then makes the expression \eqref{PRO} ill-defined. It is these divergences which prevent \eqref{PRO} from defining a \emph{bona fide} topological invariant, and cramp any connection with Ray-Singer torsion.%\footnote{Not to mention the Turaev-Viro and Crane-Yetter invariants, which are obtained by replacing $G$ with a quantum group; note however that their definition includes the number of $3$- and $4$-cells of $K$ respectively, and not just its $2$-skeleton.}

\paragraph{Cellular quantization.}

Suppose now that $d\geq3$. Let $\calA=G^{E}$ denote the space of discrete connections on $K$, and $\calF$ the subspace of \emph{flat} discrete connections, namely those for which $H_{f}=1$ for all faces $f$. In the neighborhood of $\calF$, a discrete connection $A$ can be seen as an element $(\phi,a_{\phi})\in\calF\times N_{\phi}\calF$ of the normal bundle to $\calF$, according to $A=\exp_{\phi}(a_{\phi})$.\footnote{We disregard the possibility that $\calF$ may contain singularities; see \cite{Bonzom:2010uq} for a discussion of this issue.} Furthermore, the holonomy can be expanded as
\be
H_{f}=(dH_{f})_{\phi}(a_{\phi})+\mathcal{O}(a_{\phi}^{2}).
\ee
and the Haar measure $dA$ on $\calA$ splits as
\be
dA=d\phi\, da_{\phi},
\ee
where $d\phi$ is the induced Riemannian measure on $\calF$ and $da_{\phi}$ is the Lebesgue measure on the fiber $N_{\phi}\calF$ normal to $\calF$. Finally we have
\be
\delta(H_{f})=\int_{\alg}db_{f}\ e^{i\langle b_{f},(dH_{f})_{\phi}(a_{\phi})\rangle}.
\ee
Hence \eqref{discretepartition} can be rewritten as $\int_{\calF}d\phi\ z_{\phi}$, where $z_{\phi}$ has the BF-like form
\be\label{bfpart}
z_{\phi}=\int_{N_{\phi}\calF}da_{\phi}\int_{\alg^{F}}db\ e^{is(b,a_{\phi})}
\ee
where $b=(b_{f})_{f}$ and
\be\label{actionBFdiscrete}
s(b,a_{\phi})=\sum_{f}\langle b_{f},(dH_{f})_{\phi}(a_{\phi})\rangle.
\ee

To proceed with the quantization of discrete BF theory, we must now identify the gauge symmetries of \eqref{bfpart}. To this effect, consider the \emph{discrete twisted de Rham complex}
\be\label{discretetwistedcomplex}
0\rightarrow \alg^{c_{0}}\xrightarrow{\delta_{\phi}^{0}}\dots\xrightarrow{\delta_{\phi}^{d-1}}\alg^{c_{d}}\rightarrow0.
\ee
where $c_{k}$ is the number of $k$-cells of $K$. The cochain space $\alg^{c_{k}}$ is the discrete analogue of $\Omega^{k}(M,\alg)$, and $\delta_{\phi}^{k}$ is the discrete covariant exterior derivative defined in \cite{Barrett:2009ys,springerlink:10.1007/s00023-011-0127-y}, satisfying $\delta_{\phi}^{k+1}\circ\delta_{\phi}^{k}=0$. In particular, if $\mu$ is the Maurer-Cartan form on $G$ and $a=\mu(a_{\phi})$, then $\delta^{1}_{\phi}(a)=dH_{\phi}(a_{\phi})$. Using the bracket in $\alg$, we can also consider the adjoint maps $\pp_{k}^{\phi}=(\delta_{\phi}^{k-1})^{\dagger}$, defining the dual complex to \eqref{discretetwistedcomplex}, namely
\be\label{discretetwistedcomplexdual}
0\rightarrow \alg^{c_{d}}\xrightarrow{\pp_{d}^{\phi}}\dots\xrightarrow{\pp^{\phi}_{1}}\alg^{c_{0}}\rightarrow0.
\ee

Thanks to this cohomological structure, it is easy to identify the gauge symmetries of \eqref{actionBFdiscrete}: it is simply
\be
b\mapsto b+\pp_{3}^{\phi}(X_{3})
\ee
with $X_{3}\in\alg^{c_{3}}$. Indeed, we have
\be
\langle\pp_{3}^{\phi}(X_{3}),dH_{\phi}(a_{\phi})\rangle=\langle X_{3},\delta_{\phi}^{2}\circ\delta^{1}_{\phi}(a)\rangle=0.
\ee
This is nothing but the discrete counterpart of the shift symmetry \eqref{shiftsymmetry}. When $d\geq4$, this symmetry is reducible, as $\im\pp_{4}^{\phi}\subset\ker\pp_{3}^{\phi}$, etc. That is, just as in the continuum, the reducible symmetries of the action \eqref{actionBFdiscrete} involves all the chain groups in \eqref{discretetwistedcomplexdual}, hence cells \emph{of all dimensions}.

Let us now use the resolvent method to gauge-fix the discrete shift symmetry. Assume that \eqref{discretetwistedcomplex} and \eqref{discretetwistedcomplexdual} are acyclic,\footnote{In the case where the complex \eqref{discretetwistedcomplex} is not acyclic, and in particular when the moduli space of flat connections is not discrete, this method can be amended along the lines of \cite{Adams}. This yields a similar result, except for a few more determinants.} so that $\im \pp_{3}^{\phi}$ exhausts the kernel of \eqref{actionBFdiscrete}. Then the goal is to restrict the integral over $b$ in \eqref{bfpart} to an integral over $(\im\pp_{3}^{\phi})^{\perp}$. Write formally
\be\label{divergent}
\int_{\alg^{F}}db\ e^{is(b,a_{\phi})}=\Vol(\im\pp_{3}^{\phi})\int_{(\im\pp_{3}^{\phi})^{\perp}}db_{\perp}\ e^{is(b,a_{\phi})}
\ee
and observe that, since $\pp_{3}^{\phi}$ provides an isomorphism between $\alg^{c_{3}}/\ker \pp_{3}^{\phi}=\alg^{c_{3}}/\im\pp_{4}^{\phi}$ and $\im\pp_{3}^{\phi}$, and moreover $\delta_{\phi}^{2}=(\pp_{3}^{\phi})^{\dagger}$, we can write
\be
\Vol(\im\pp_{3}^{\phi})=\det (\delta_{\phi}^{2}\pp_{3}^{\phi})^{1/2}\f{\Vol(\alg^{c_{3}})}{\Vol(\im \pp_{4}^{\phi})}.
\ee
Iterating this recursive relation, we get
\be
\Vol(\im \pp_{3}^{\phi})=\prod_{j=2}^{d-1}\det (\delta_{\phi}^{j}\pp_{j+1}^{\phi})^{(-1)^{j}/2}\Vol(\alg^{c_{j}})^{(-1)^{j}}.
\ee
Now, let us \emph{pretend} that the chain spaces $\alg^{c_{j}}$ have unit volume: this is the meaning of the expression ``dividing by an infinite volume'' underlying the gauge-fixing procedure. (Precisely the same step is taken in the continuum quantization of BF theory.) Then we can replace \eqref{divergent} by the finite quantity
\be
\prod_{j=2}^{d-1}\det(\delta_\phi^j\pp_{j+1}^\phi)^{(-1)^j/2}\int_{(\ker\pp_{3}^{\phi})^{\perp}}db_{\perp}e^{is(b_{\perp},a_{\phi})}.
\ee
Hence, returning to \eqref{bfpart} and performing the integral over $b_{\perp}$, we get as the definition of gauge-fixed version of $z_{\phi}$
\be
z'_{\phi}= \prod_{j=2}^{d-1}\det(\delta_\phi^j\pp_{j+1}^\phi)^{(-1)^j/2}\int_{N_{\phi}\calF}da_{\phi}\,\delta\big(dH_{\phi}(a_{\phi})\big).
\ee
The remaining integral over $a_{\phi}$ is now well-defined and gives $\det(\delta^{1}_{\phi}\pp_{2}^{\phi})^{-1/2}$. Hence we obtain for the gauge-fixed partition function $Z'_{\textrm{BF}}=\int_{\calF}d\phi\,z'_{\phi}$:
\be
Z'_{\textrm{BF}}=\int_{\calF}d\phi\, \prod_{j=1}^{d-1}\det(\delta_\phi^j\pp_{j+1}^\phi)^{(-1)^j/2}.
\ee
The integral over $\calF$ can be pulled back to to the moduli space of flat discrete connections $\calM=\calF/G^{c_{0}}$ by integrating along the gauge orbits of each flat connection \cite{Barrett:2009ys,Bonzom:2010uq}. This yields one more determinant $\det(\delta^{0}_{\phi}\pp_{1}^{\phi})^{1/2}$, and thus
%\be
%Z'_{\textrm{BF}}=\sum_{[\phi]\in\calM}\,\prod_{j=0}^{d-1}\det(\delta_\phi^j\pp_{j+1}^\phi)^{(-1)^j/2},
%\ee
%i.e.
\be\label{discreteresult}
Z'_{\textrm{BF}}=\sum_{[\phi]\in\calM}\Tor_{[\phi]}
\ee
with
\be
\Tor_{[\phi]}=\prod_{j=0}^{d-1}\det(\pp_{j+1}^\phi\delta_\phi^j)^{(-1)^j/2}.
\ee

The expression \eqref{discreteresult} is a topological invariant of $K$. In particular, the quantity $\Tor_{[\phi]}(K,G)$ is the twisted Reidemeister torsion, which is known to coincide with the twisted analytic torsion. Thus, \eqref{discreteresult} matches with the continuum result, consistently with the general expectation that, for a TQFT with finitely many degrees of freedom, discretization should play no physical r\^ole.

%The cochain space $\alg^{c_{k}}$ is the discrete analogue of $\Omega^{k}(M,\alg)$. To describe the corresponding analogue of the covariant exterior derivative, first pick a reference vertex $v^i_\alpha$ on the boundary of each $i$-cell $e^i_\alpha$ of $K$. Then choose for each $(i+1)$-cell $e^{i+1}_{\beta}$ adjacent to $e^i_\alpha$ an edge-path connecting $v^i_{\alpha}$ and $v^{i+1}_{\beta}$ on the boundary of $e^{i+1}_{\beta}$, and form the parallel transport operator $P_{\phi}(v^i_{\alpha},v^{i+1}_{\beta})$ associated to this edge-path.

%It is easy to see that $s(a,b)=\langle b,dH(a)\rangle$ is invariant under the \emph{discrete shift symmetry}
%\be
%b\mapsto b+dH_{\phi}(c).
%\ee
%This transformation is stricly analogous to the continuum shift symmetry; in particular, if $d\geq4$, it is reducible.

%%%%%%%%%%%%%%%%%%%
\paragraph{Relation to the loop formalism.}
%%%%%%%%%%%%%%%%%%%

The above method naturally gives rise to the loop quantization of BF theory. In the loop approach, one quantizes before restricting to flat gauge fields. Given an embedded, closed graph $\gamma$, \emph{cylindrical} wave functions are functions of the Wilson lines along the lines of $\gamma$. For each graph there is a Hilbert space whose measure is given by the Haar measure of $G$ on each line, $\prod_e dg_e$. The Hilbert spaces of two different graphs are orthogonal. The standard gauge symmetry requires invariance under $G$-translation on the source and end nodes of the lines.

Heuristically, the transition amplitudes in the continuum \eqref{transition} suggest that they can be formulated in the loop approach by taking as boundary states cylindrical functions restricted to the moduli space $\calM$, the torsion still providing the measure. Assume $M$ has two disconnected boundaries $N_1, N_2$, with two closed, embedded graphs $\gamma_1, \gamma_2$ associated with two cylindrical functions $\Psi_{\gamma_1}, \Psi_{\gamma_2}$. The transition is regularized by choosing a cell decomposition $K$ of $M$ such that $\gamma_1, \gamma_2$ are included into the 1-skeleton. The ungauge-fixed transition amplitude reads
\be
\langle \Psi_{\gamma_2}\vert Z_{\rm BF}\vert \Psi_{\gamma_1}\rangle = \int \prod_{e} dg_e\ \Psi^*_{\gamma_2}(g_e) \Psi_{\gamma_1}(g_e) \prod_f \delta(H_f).
\ee
As the shift symmetry does not act on Wilson lines, the process of the previous section applies. The wave-functions are evaluated on $\calM$ because there are no fluctuations around flat connections, yielding
\be \label{looptransition}
\langle \Psi_{\gamma_2}\vert Z'_{\rm BF}\vert \Psi_{\gamma_1}\rangle = \sum_{[\phi]\in\calM} \Psi_{\gamma_2}^*([\phi])\,\Tor_{[\phi]}\,\Psi_{\gamma_1}([\phi]).
\ee
Finally, the regulator $K$ can be removed thanks to the topological invariance of the torsion, which makes the continuum limit result into the above formula. Let us mention an outcome of this result: the loop quantization of the BF model does not distinguish knottings of the graphs $\gamma_{1,2}$.

%\paragraph{On the diffeomorphism symmetry}

\paragraph{Conclusion.}

We have performed a topological quantization of discrete BF theory, proving its equivalence to the usual quantization in the continuum. This result solves several open problems of the field and extends previous results obtained in dimension 3 to arbitrary dimensions: $(1)$ transition amplitudes are finite, answering the issue of bubble divergences \cite{Freidel2002, Bonzom:2010uq}; $(2)$ the gauge symmetries in the discrete setting exist, generalizing \cite{Freidel2002, Freidel2004}, and $(3)$ they can be gauge-fixed to derive the loop quantization, generalizing \cite{noui-perez}; $(4)$ as a result, one gets a topological invariant, which proves that the classical gauge symmetries are correctly promoted to the quantum level.

The crucial steps of our quantization require to take into account cells of all dimensions in the cell complex, and not just its $2$-skeleton like in the ``spinfoam quantization''. A challenge for future investigations is to find a representation of \eqref{looptransition} as a state-sum, as is done in the latter approach.\footnote{This is of direct relevance for non-trivial topologies. But in the spinfoam literature, one is mainly interested in the local degrees of freedom and not topological ones, so it is usually assumed that one can work safely on spheres for which there is no difficulties in gauge-fixing. We thank A. Perez for pointing this out to us.}

The last issue we mentioned in the introduction is the major difficulty in quantum gravity: understanding the quantum version of diffeomorphism-invariance. It is well-known that diffeomorphism-invariance in the BF model is contained within its shift symmetry \cite{Horowitz1989a}. Hence the substance of general relativity is to break the topological invariance while preserving diffeomorphism-invariance. Spinfoam models for quantum gravity are very much in line with this idea, as they start by quantizing BF theory and then introduce some breaking of the shift symmetry to restore the local degrees of freedom. It is also known that discrete models of gravity generically break diffeomorphism-invariance \cite{Dittrich:2012qb}. Showing that it is restored in the continuum limit (after some coarse-graining, or summing over spinfoams appropriately) is one of the main programs in the spinfoam approach. Now that the shift symmetry is correctly controlled in the discrete setting, we feel that the noose is tightening around diffeomorphisms.

\smallskip 
We are glad to thank Carlo Rovelli, Alejandro Perez and Simone Speziale for their critical reading of an earlier version of this manuscript, as well as Bianca Dittrich, Razvan Gurau and Aristide Baratin for numerous discussions on the invariance of spinfoam amplitudes. Research at Perimeter Institute is supported by the Government of Canada through Industry Canada and by the Province of Ontario through the Ministry of Research and Innovation.

\bibliographystyle{utcaps}

\begin{thebibliography}{10}

\bibitem{Baez1994a}
J.~C. Baez, ``{Strings, loops, knots and gauge fields},'' in {\em Knots and
  Quantum Gravity}, J.~Baez, ed.
\newblock Oxford University Press, 1994.

\bibitem{Reisenberger1994}
M.~Reisenberger, ``{Worldsheet formulations of gauge theories and gravity},''
  \href{http://arxiv.org/abs/9412035}{{\ttfamily arXiv:9412035 [gr-qc]}}.

\bibitem{Reisenberger1996}
M.~P. Reisenberger and C.~Rovelli, ``{"Sum over Surfaces'' form of Loop Quantum
  Gravity},'' {\em Phys. Rev. D} {\bfseries 56} no.~6, (1996) 3490--3508,
  \href{http://arxiv.org/abs/9612035}{{\ttfamily arXiv:9612035 [gr-qc]}}.

\bibitem{Freidel1998a}
L.~Freidel and K.~Krasnov, ``{Spin foam models and the classical action
  principle},'' {\em Adv. Theor. Phys.} {\bfseries 2} (1998) 1221--1285,
  \href{http://arxiv.org/abs/9807092}{{\ttfamily arXiv:9807092 [hep-th]}}.

\bibitem{Baez:1999sr}
J.~C. Baez, ``{An introduction to spin foam models of BF theory and quantum
  gravity},'' {\em Lect. Notes Phys.} {\bfseries 543} (2000) 25--94,
  \href{http://arxiv.org/abs/9905087}{{\ttfamily arXiv:9905087 [gr-qc]}}.

\bibitem{PR}
G.~Ponzano and T.~Regge, ``{Semiclassical limit of Racah coefficients},'' in
  {\em Spectroscopic and group theoretical methods in physics}.
\newblock North-Holland, New York, 1968.

\bibitem{Ooguri:1992eb}
H.~Ooguri, ``{Topological lattice models in four-dimensions},''
 % \href{http://dx.doi.org/10.1142/S0217732392004171}
  {\em Mod. Phys. Lett.} {\bfseries A7} (1992) 2799--2810,
  \href{http://arxiv.org/abs/9205090}{{\ttfamily arXiv:9205090 [hep-th]}}.

\bibitem{Engle:2007uq}
  J.~Engle, R.~Pereira and C.~Rovelli,
  ``The Loop-quantum-gravity vertex-amplitude,''
  {\em Phys.\ Rev.\ Lett.}\  {\bfseries 99} (2007) 161301
  \href{http://arxiv.org/abs/0705.2388}{{\ttfamily arXiv:0705.2388 [gr-qc]}}.
  %%CITATION = ARXIV:0705.2388;%%

\bibitem{Perez:2003vx}
A.~Perez, ``{Spin foam models for quantum gravity},'' {\em Class. Quant. Grav.}
  {\bfseries 20} (2003) R43, \href{http://arxiv.org/abs/0301113}{{\ttfamily
  arXiv:0301113 [gr-qc]}}.

\bibitem{Rovelli2011c}
C.~Rovelli, ``{Zakopane lectures on loop gravity},''
  \href{http://arxiv.org/abs/1102.3660 [gr-qc]}{{\ttfamily arXiv:1102.3660
  [gr-qc]}}.
  
  
\bibitem{Freidel2002}
L.~Freidel and D.~Louapre, ``{Diffeomorphisms and spin foam models},'' {\em
  Nucl. Phys. B} {\bfseries 662} no.~1-2, (2002) 19,
  \href{http://arxiv.org/abs/0212001}{{\ttfamily arXiv:0212001 [gr-qc]}}.

\bibitem{Freidel2004}
L.~Freidel and D.~Louapre, ``{Ponzano-Regge model revisited I: Gauge fixing,
  observables and interacting spinning particles},'' {\em Class. Quant. Grav.}
  {\bfseries 21} no.~24, (2004) 48,
  \href{http://arxiv.org/abs/0401076}{{\ttfamily arXiv:0401076 [hep-th]}}.

\bibitem{noui-perez}
  K.~Noui and A.~Perez,
  ``Three-dimensional loop quantum gravity: Physical scalar product and spin
  foam models,''
  {\em Class.\ Quant.\ Grav.}\  {\bfseries 22}, 1739 (2005)
  \href{http://ariv.org/abs/0402110}{{\ttfamily arXiv:0402110 [gr-qc]}}.
  %%CITATION = CQGRD,22,1739;%%

\bibitem{Barrett:2009ys}
J.~W. Barrett and I.~Naish-Guzman, ``{The Ponzano-Regge model},'' {\em Class.
  Quant. Grav.} {\bfseries 26} (2009) 155014,
  \href{http://arxiv.org/abs/0803.3319}{{\ttfamily arXiv:0803.3319}}.

\bibitem{Baratin:2010wi} 
  A.~Baratin and D.~Oriti,
  ``Group field theory with non-commutative metric variables,''
  {\em Phys.\ Rev.\ Lett.}\  {\bfseries 105}, 221302 (2010)
  \href{http://ariv.org/abs/1002.4723}{{\ttfamily arXiv:1002:4723 [hep-th]}}.
  %%CITATION = ARXIV:1002.4723;%%

\bibitem{Baratin:2011tg} 
  A.~Baratin, F.~Girelli and D.~Oriti,
  ``Diffeomorphisms in group field theories,''
  {\em Phys.\ Rev.\ D} {\bfseries 83}, 104051 (2011)
  \href{http://arxiv.org/abs/1101.0590}{{\ttfamily arXiv:1101.0590 [hep-th]}}.
  %%CITATION = ARXIV:1101.0590;%%

\bibitem{Dittrich:2012qb} 
  B.~Dittrich,
  ``How to construct diffeomorphism symmetry on the lattice,''
  \href{http://arxiv.org/abs/1201.3840}{{\ttfamily arXiv:1201.3840 [gr-qc]}}.
  %%CITATION = ARXIV:1201.3840;%%

\bibitem{perezrovelli}
V.~Bonzom and M.~Smerlak,
 ``Cellular quantization of discrete BF theory'',
 {\em to appear}.

\bibitem{discreteBF}
A.~Perez and C.~Rovelli,
 ``Perturbative Finiteness in Spin-Foam Quantum Gravity'',
 {\em Phys. Rev. Lett}{\bfseries 87}, 181301 (2001).

\bibitem{Horowitz1989a}
G.~T. Horowitz, ``{Exactly soluble diffeomorphism invariant theories},'' {\em
  Comm. Math. Phys.} {\bfseries 125} no.~3, (1989) 417--437.

\bibitem{Blau1991}
M.~Blau and G.~Thompson, ``{Topological gauge theories of antisymmetric tensor
  fields},'' \href{http://dx.doi.org/10.1016/0003-4916(91)90240-9}{{\em Ann.
  Phys.} {\bfseries 205} no.~1, (Jan., 1991) 130--172}.

\bibitem{blau-thompson-torsion}
  M.~Blau and G.~Thompson,
  ``A new class of topological field theories and the Ray-Singer torsion,''
  {\em Phys.\ Lett.\ B} {\bfseries 228}, 64 (1989).
  %%CITATION = PHLTA,B228,64;%%

\bibitem{Schwarz1978}
A.~S. Schwarz, ``{The partition function of degenerate quadratic functional and
  Ray-Singer invariants},'' \href{http://dx.doi.org/10.1007/BF00406412}{{\em
  Lett. Math. Phys.} {\bfseries 2} no.~3, (Jan., 1978) 247--252}.

\bibitem{Adams}
D.~H. Adams and S.~Sen, ``{Partition Function of a Quadratic Functional and
  Semiclassical Approximation for Witten's 3-Manifold Invariant},''
  \href{http://arxiv.org/abs/9503095}{{\ttfamily arXiv:9503095 [hep-th]}}.

\bibitem{Perez:2000fs}
A.~Perez and C.~Rovelli, ``{A spin foam model without bubble divergences},''
  \href{http://dx.doi.org/10.1016/S0550-3213(01)00030-X}{{\em Nucl. Phys. B}
  {\bfseries 599} (2001) 255--282},
  \href{http://arxiv.org/abs/0006107}{{\ttfamily arXiv:0006107 [gr-qc]}}.
\bibitem{springerlink:10.1007/s00023-011-0127-y}
V.~Bonzom and M.~Smerlak, ``{Bubble Divergences: Sorting out Topology from Cell
  Structure},'' {\em Annales Henri Poincare} (2011) 1--24,
  \href{http://arxiv.org/abs/1103.3961}{{\ttfamily arXiv:1103.3961}}.

\bibitem{Witten1991a}
E.~Witten, ``{On quantum gauge theories in two dimensions},'' {\em Comm. Math.
  Phys.} {\bfseries 141} no.~1, (1991) 153--209.

\bibitem{Bonzom:2010uq}
V.~Bonzom and M.~Smerlak, ``{Bubble divergences from twisted cohomology},''
  {\em to appear in Comm. Math. Phys.} ,
  \href{http://arxiv.org/abs/1008.1476}{{\ttfamily arXiv:1008.1476}}.

\end{thebibliography}
\providecommand{\href}[2]{#2}\begingroup\raggedright\endgroup

\end{document}